# Fermi arcs and the nature of carriers in cuprate HTSC

.K.V. Mitsen [1] & O.M. Ivanenko

Lebedev Physical Institute RAS, 119991 Moscow, Russia

**The nature of the normal state and the mechanism of superconductivity in HTSC cuprates remain under discussion until now. However, a number of experimental facts may be considered firmly established: 1) There is a *d*-wave superconducting energy gap on the 2D Fermi surface (FS) of optimally doped HTSC cuprates, with zeros in node directions [1]. 2) The transition to normal state is accompanied by the closing of the gap across parts of the FS and the formation of Fermi arcs centered at the nodes. Meanwhile, the gap in antinode directions virtually does not change across $T_c$ ("pseudogap") and closes at some temperature $T^*$ well above $T_c$ [2,3]. 3) As the doping level is reduced, the pseudogap increases and deviates from the simple *d*-wave behavior [4]. In deeply underdoped samples, Cooper pairs form islands in *k*-space around the nodal regions [5]. 4) Above $T_c$, underdoped samples exhibit giant Nernst effect and anomalous diamagnetism [6,7]. Here we suggest a simple physical model of HTSC cuprates that provides explanation of the features listed above. According to this model, the unusual properties of these compounds result from their unique electronic structure favourable for the formation of diatomic negative-U centers (NUCs) and realization of an unusual mechanism of electron–electron interaction.**

The suggested model is based on the mechanism of the formation of NUCs in HTSCs that we proposed previously [8]. Its essence can be understood from Fig. 1a, which shows the electronic spectrum of a $CuO_2$ plane of an undoped HTSC. It is known that the electronic structure of the insulating phase of HTSCs in the vicinity of the Fermi energy $E_F$ is described well by the model of an insulator with a charge-transfer gap [9]. In this scheme, the energy of the lowest-lying excitation $\Delta_{ct}$ (~2 eV) is related to the transfer of an electron from an oxygen ion to a neighboring copper ion, $3d^9L \rightarrow 3d^{10}L^-$ (here, $3d^9L$ denotes the state with a hole on the 3d-shall of Cu ion and a completely occupied 2p-shell of the neighboring ligand (oxygen), and $3d^{10}L^-$ is the state with a completely occupied 3d-shell of Cu and a hole on the ligand ion). Note that the $Cu3d^{10}L$ state is considerably higher in energy than $Cu3d^{10}L^-$.

---

[1] -mail address: mitsen@sci.lebedev.ru

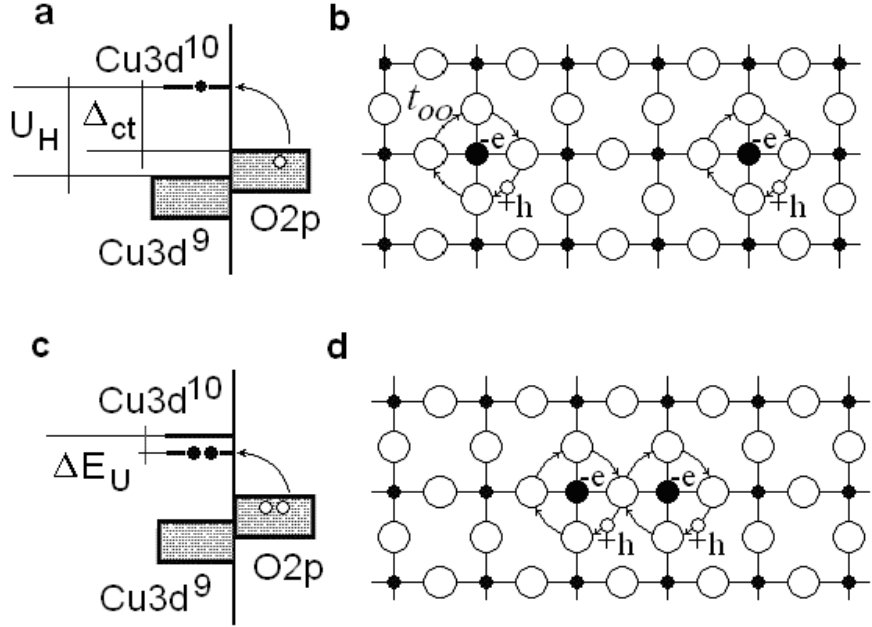

Fig. 1. (a) The electronic spectrum of an undoped $CuO_2$ plane; $U_H$ is the energy of repulsion between two electrons on a Cu ion. The $\Delta_{ct}$ gap for the lowest-energy excitation corresponds to the transfer of an electron from O to the nearest Cu ion with the formation of a hole distributed over four surrounding O ions (b). (c) The energy of two such quasi-atomic excitations can be decreased by $\Delta E_U$ if they arrange side by side and form a quasi-molecule (d).

The hole $L^-$ resulting from the electron transfer is spread over the four surrounding oxygen ions (Fig. 1b) due to an overlap between nearest-neighbor oxygen orbitals ($t_{OO}$ is the hopping integral between the $p\sigma$ orbitals of the nearest O ions). This exciton-like state (an electron on copper and a hole on the surrounding oxygen ions) resembles the hydrogen atom. Pushing this analogy further, we can suggest that the energy of a state with two excitations of this kind is lower (Fig. 1c) if two such quasi-atoms neighbor one another forming a quasi-molecule (Fig. 1d). This is possible owing to the formation of a bound state (of Heitler–London type) of two electrons on neighbouring Cu ions and two holes that appear in the vicinity of that pair of ions. A bound state appears for the bonding orbital of the hole pair owing to the possibility for each of the holes to be located between Cu ions and to be attracted to both electrons on these ions simultaneously.

Thus, we argue that diatomic NUCs, characterized by negative electron-correlation energy, form with the participation of pairs of neighboring Cu ions in $CuO_2$ planes (Fig. 1c). An estimation of the binding energy yields $\Delta E_U \sim 0.2$–$0.3$ eV [8].

Now, if $\Delta_{ct}$ is reduced somehow until transitions between $3d^{10}L^-$ and $3d^9L$ states become possible, a half-filled band with the FS shown in Fig. 2 will arise [10].

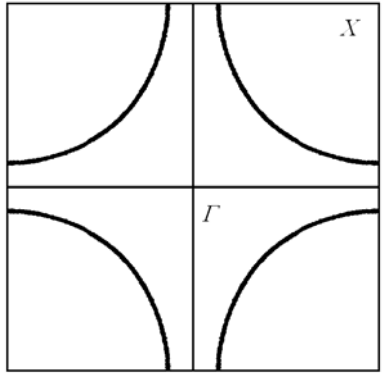

Fig. 2. The FS of the CuO$_2$ plane, calculated in the framework of a tight-binding approximation taking into account the nearest-neighbour (i.e., O–Cu) and the next-nearest-neighbour (i.e., O–O) interaction. [10]

**The role of doping**

What factors might possibly lead to a decrease in $\Delta_{ct}$ so that hybridization between $3d^{10}L^-$ and $3d^9L$ states takes place and an active NUC, which band electrons can interact with, forms on a given pair of Cu ions in the CuO$_2$ plane? The simplest situation of this kind occurs when, in the immediate vicinity of each of the Cu cations from the pair, there is a positive charge which reduces the energy of the $3d^{10}L^-$ state of the corresponding cation by the value required. This situation, as we will see below, does take place in HTSCs upon doping.

It is well known that mobile carriers in HTSC cuprates appear as a result of doping, and it is generally believed that the charge carriers arising under doping are directly transferred to CuO$_2$ planes from the dopants. In contrast, we suppose that charges introduced upon doping remain localized in the vicinity of the dopant ions [11,12] and their role consists in a local modification of the electron structure of nearby CuO$_2$ planes, where the energy of the upper Hubbard Cu$3d^{10}$ state of the original (undoped) insulator is lowered and the valence-band states hybridize with the low-lying Cu$3d^{10}L^-$ state.

As an example, consider YBa$_2$Cu$_3$O$_{6+\delta}$, whose structural fragments are outlined in Fig. 3[2]. In this case, doping is carried out by adding an excess amount of oxygen $\delta$ to the chains of insulating YBa$_2$Cu$_3$O$_6$. We presume that holes introduced upon doping remain localized in the O sheets formed around Cu ions in the chains. In the case where three consecutive positions in a chain are occupied by O ions, each of the two resulting O sheets contains a hole distributed over the four O ions (⊕) in a given sheet. It is of $\approx +e/4$ at the apical O ions (⊕), closest to the Cu ions of CuO$_2$ planes, that affect most profoundly the Cu-ion energy levels, lowering the energy of Cu$3d^{10}L^-$ states. Inasmuch as screening exists on a scale larger than the interatomic distances but is absent at smaller distances, it can be assumed that these charges are virtually unscreened.

---

[2] The mechanism of doping in La$_{2-x}$Sr$_x$CuO$_4$ and the explanation of its phase diagram are developed in [8].

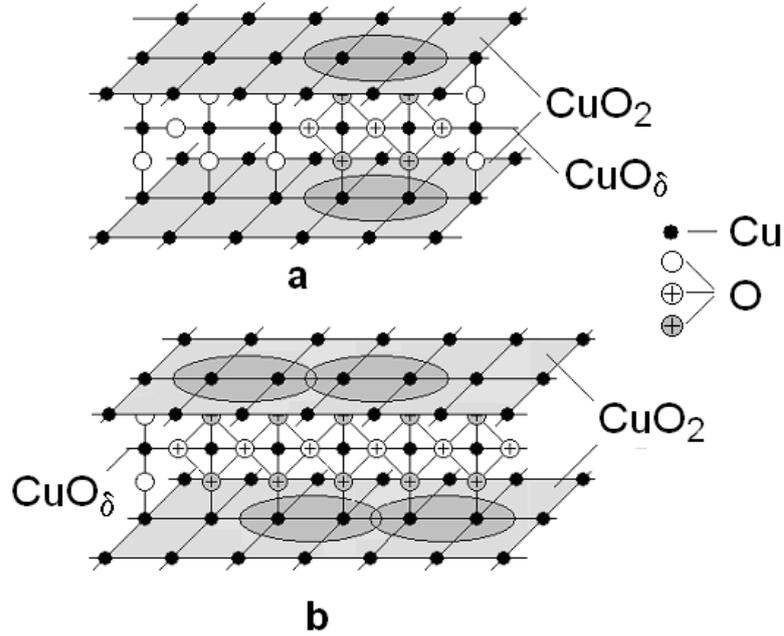

Fig. 3. (a) In $YBa_2Cu_3O_{6+\delta}$, a NUC (ellipse) is formed by a pair of Cu ions in the $CuO_2$ plane in the presence of three successively filled oxygen positions in a $CuO_3$ chain over (under) this pair of Cu ions. (b) The formation of linear NUC clusters in the $CuO_2$ planes in $YBa_2Cu_3O_{6+\delta}$ by the sequence of oxygen ions in chains. Oxygen ions in the $CuO_2$ planes are not shown.

Taking only the nearest-neighbor interaction into account, one can estimate the reduction in the energy of the $3d^{10}L^-$ state of the nearest Cu ion in the $CuO_2$ plane caused by the doping as $\Delta E = e^2/4r \approx 1.8–1.9$ eV $\sim \Delta_{ct}$ (here, $r \approx 0.2$ nm is the spacing between an apical O ion and the nearest Cu ion). An energy lowering of this order is sufficient for hybridization of $3d^{10}L^-$ and $3d^9L$ states to occur and an active NUC to form on the given pair of Cu ions. The limiting value $\delta = 1$ for $YBa_2Cu_3O_{6+\delta}$ corresponds to the case where O positions in the chains are filled completely and each Cu ion in a $CuO_2$ plane belongs to an NUC - in other words, the percolation cluster of NUCs occupies the entire $CuO_2$ plane.

With decreasing $\delta$, the capacity of the percolation cluster decreases and, for $\delta < 0.8$ [8], the $CuO_2$ plane breaks into finite NUC clusters.

**The charge-transport processes. Superconductivity and the mechanism of the hole carrier generation**

Thus, under a proper doping, we obtain a half-filled band formed by $3d^{10}L^-$ and $3d^9L$ states with one electron and one hole per $CuO_2$ cell. Evidently, a material with such an electronic structure should exhibit unusual properties. On the one hand, it possesses a ungaped FS similarly to metals, since nothing prevents electrons from "leaking" in the momentum space so that their energy varies and at that each $CuO_2$ cell always contains a single electron. On the other hand, such a material is an insulator (at $T = 0$), since each cell can be occupied by only one electron and incoherent single-electron transport is not possible.

At the same time, transport processes other than incoherent single-electron transport can occur in such a system, which are coherent electron transport (where the electron condensate moves as a whole) and incoherent hole transport (if there exists a mechanism providing for the mobile hole generation).

Let us consider possible mechanisms, which can be responsible for these transport processes. First, we study the possibility of establishing a superconducting coherent state. Let us begin with the case where each Cu ion belongs to some NUC. This corresponds to the situation of the optimal doping, with the entire $CuO_2$ plane representing a single NUC cluster with a common pair level. Then, owing to the virtual transitions of electron pairs to NUCs, states ($k\uparrow,-k\downarrow$) in the vicinity of the FS are coupled, that should provide for superconducting pairing in the system. The pairing potential $\Delta$ will exhibit a pronounced $k$ dependence, vanishing in the nodal directions (i.e., along the directions of O–O bonds) and attaining maxima in the antinodal directions (i.e., along the directions of Cu–O bonds), where the rate of pair transitions to NUCs is maximum. Thus, $\Delta(k)$ has a $d$-wave character. For some temperature $T = T_c$, determined by $|\Delta(k)|$, the electron system performs a transition to the superconducting state, which ensures the charge transport for $T < T_c$.

Another important aspect of the system under consideration is the mechanism of generation of quasiparticle excitations at $T < T_c$. Apart from ordinary thermal excitations, whose spectrum is described by Bogolyubov dispersion curves (Fig. 4a), a special type of two-particle excitations is possible. They appear owing to the pair hybridization of the NUC level with the band states. The pair hybridization results in broadening of the pair level, with the width $\Gamma$ depending on the temperature [13,14]:

$$\Gamma \approx kT \cdot (V/E_F)^2 \qquad (1)$$

(here, $V$~0.5 eV is the one-particle hybridization constant, $E_F$~0.2–0.3 eV is the Fermi energy [15-17], and $T$ is the temperature). From this expression we obtain $\Gamma$~(3÷5) $kT$.

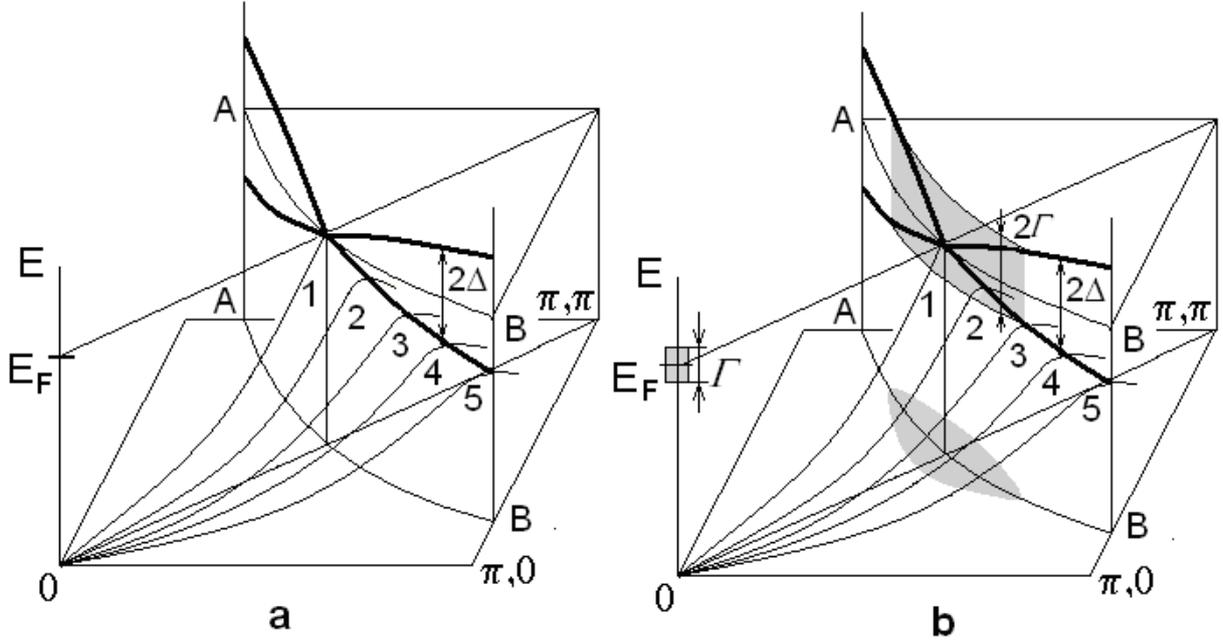

Fig. 4. Development of Fermi arcs in cuprate HTSCs as a result of interaction of band electrons with NUCs: a) $T = 0$, b) $T > 0$. Curves $01$–$05$ are lower branches of the Bogolyubov quasiparticle dispersion curves; the upper branches are not shown. $AB$ is the Fermi contour; $\Gamma$ is the width of NUC pair level. The shaded area around $(\pi/2; \pi/2)$ is the region of momenta of electronic pairs $(k_1, k_2)$ available for transitions to NUCs at a given temperature. The solid curve (1–5) is the locus of extremum points of lower Bogolyubov dispersion branches. The upper solid curve is the locus of extremum points of upper Bogolyubov dispersion branches.

The pair hybridization results in transitions of electron pairs $(k_1, k_2)$ to NUCs. These transitions are accompanied by the appearance of two quasiparticles $-k_1, -k_2$ satisfying the condition $E(k_1) + E(k_2) < \Gamma$, where the energies $E(k_1)$ and $E(k_2)$ are measured from Fermi level. Equality $E(k_1) + E(k_2) = \Gamma$ takes place, when two electrons from different two pairs transit to NUC, having energy-$\Gamma/2$ everyone, and give rise to 2 quasiparticles with total energy $\Gamma$.

As the temperature increases, the region of energies $E$ for which real transitions of electron pairs to NUCs are possible stretches from point $(\pi/2; \pi/2)$ along the direction of the "crest" of the dispersion, so that a "belt" of height $2\Gamma$, thickness $\Delta k(k)$, and length $L$ along the contour of the FS is formed (Fig. 4b). The arc length $L(T)$ is determined by the condition $\Gamma(T) = \Delta(k)$. The number of such states increases with the temperature as $T^2$ (the hatched area around $(\pi/2; \pi/2)$ in Fig. 4b).

The NUC occupancy $\eta$ ($0 < \eta < 2$) is determined by the condition that rates of transitions between the band and the pair-level states in both directions are equal. According to (1), the rate of pair level to the band transitions $\eta\Gamma \propto T\eta$. The rate of the reverse process is determined by the number of band states from which transitions to NUCs is possible and the number of empty NUCs, which means this rate is proportional to $T^2(2-\eta)$. Thus,

$$\eta = 2T/(T+T_0) \qquad (2)$$

where constant $T_0$ is independent of the temperature.

So, transitions of electron pairs ($k_1,k_2$) to NUCs are accompanied by depairing and result in the formation of Bogolyubov quasiparticles [18] within a belt of length $L(T)$ and height $2\Gamma(T)$. These processes should lead to vanishing of the superconducting order parameter around nodes in a arc of length $L(T)$ along Fermi contour. However, owing to the preservation of coherence in the system, a nonzero order parameter persists on the entire FS excluding the nodes. At the same time, filling of NUCs with real electrons leads to a reduction in the number of NUCs available for virtual transitions of electron pairs. As the temperature increases, the NUC occupancy approaches a critical value $\eta_c$ at which point the superconducting coherence is destroyed and a transition to the normal state takes place. The gap closes along an arc of length L around each nodal direction at the FS due to depairing. Meanwhile, along the remaining part of the FS, there still exists a gap (the pseudogap), which in this situation corresponds to incoherent pairing. It is exactly this behavior that is observed in ARPES [19].

Now, let us consider the mechanism of the normal-state conductivity. As we mentioned earlier, in the system under study in the normal state, each electron should be localized in its cell. On the other hand, two-particle hybridization results in a shift of the part of electron density to the localized NUC states, which results in the creation of holes in the band, mostly at the O orbitals. An overlap between the hole wave functions belonging to different NUCs leads to the formation of extended hole states providing for the transport of holes across the crystal. The number of such mobile holes per one Cu ion $n_{Cu}=\eta/2=T/(T+T_0)$. The constant $T_0$ can be determined from the Hall measurements, which yield $T_0 = 390$ K for $YBa_2Cu_3O_7$ [8].

Note that processes of hole transport (at $T > T_c$) and coherent electron transport (at $T < T_c$) will be characterized by the opposite signs of the charge carriers [20,21].

**The region of underdoping. Fluctuations and "pseudogap" anomalies**

With decreasing doping level, the number of NUCs decreases and there appear Cu ions that do not belong to NUCs. Such an ion can be thought of as a defect introducing an extra positive potential $\sim\Delta_{ct}$. In the one-dimensional problem, as shown in [22], in the presence of such defect an upper state becomes split off from the band and localized in the vicinity of the defect. In our two-dimensional case, the number of split-off states will depend on the direction of *k*. As a function of angle, the number of split-off states increases with increasing contribution from Cu orbitals; i.e., this number is the largest for states in the direction of Cu–O bonds. As the number of such defects increases, the states under FS become split off. This leads to the formation of an insulating gap over the FS region from points ($\pm\pi,0; 0,\pm\pi$) towards the nodal directions [4]. The superconducting gap persists only in the FS region adjacent to the nodes, forming islands in the *k* space [23]. Analogous result was obtained by Monte Carlo techniques in [24]. For a completely undoped (insulating) sample the contour along which an abrupt reduction in the occupation numbers $n(k)$ occurs retains the shape of the original FS contour existing in an optimally doped material (remnant Fermi surface[25]).

In the real space, a reduction in the doping level below the optimum will lead to a decrease in the size of the percolation cluster of NUCs. In $YBa_2Cu_3O_{6+\delta}$, for $\delta<0.8$ the percolation cluster breaks into finite NUC clusters whose average size decreases with the doping level [26]. In these conditions, the role of the fluctuations in the NUC occupancy increases significantly.

According to the suggested model, a transition from the superconducting to the normal state is related to the phase coherence disappearance taking place as the NUC occupancy approaches a critical value. Thus, whenever a fluctuation causes a decrease in the NUC occupancy, conditions for the restoration of superconducting coherence occur, which can result in "switching-on" of the superconductivity in the temperature range $T^*>T>T_{c\infty}$ (here, $T_{c\infty}$ is the equilibrium value of $T_c$ for an infinite NUC cluster). On the other hand, fluctuation-related increases in the NUC occupancy lead to the disruption of coherence and to "switching-off" of the superconductivity for $T_c<T<T_{c\infty}$. Large fluctuations in the NUC occupancy, corresponding to considerable deviations of $T^*$ and $T_c$ from $T_{c\infty}$, are possible in underdoped samples, where NUCs are combined into finite clusters. As the doping level is reduced, the average size of these clusters decreases and relative fluctuations in the NUC occupancy in these clusters grow (i.e., $T^*$ increases and $T_c$ decreases). Meanwhile, in an "overdoped" sample, which can be considered as a single infinite superconducting cluster, such fluctuations become impossible.

In the context of the suggested model, dependences of $T^*$ and $T_c$ on the cluster size (and on the doping level $\delta$) can be determined in the following way. We suppose that, for $\delta < \delta_c$, NUCs form finite clusters of some average size $S(\delta)$, and the sample represents a Josephson medium, where superconductivity of the entire system appears due to the Josephson coupling between superconducting clusters. We measure the size $S$ of a cluster by the number of Cu cites it contains.

Consider a cluster in the $CuO_2$ plane containing a number of NUCs. Then, according to (2), the number of electrons at NUCs in the given cluster at temperature $T$ equals $N=TS/(T+T_0)$. Owing to fluctuations, this number may vary by $\pm\sqrt{N}$. The condition for fluctuating "switch-on" ("switch-off") of superconductivity in the cluster at temperature $T^*$ ($T_c$) can be written out as $N(T)\pm\sqrt{N(T)} = N_c$, where $N_c$ is the number of electrons at NUCs in the cluster for $T=T_{c\infty}$. The plus sign corresponds to $T=T^*$, and the minus sign corresponds to $T=T_c$. Thus,

$$TS/(T+T_0)\pm(TS/(T+T_0))^{1/2}=T_{c\infty}S/(T_{c\infty}+T_0). \qquad (3)$$

Solving equations (3), we can find the dependences of $T^*$ and $T_c$ for YBCO (with $T_{c\infty}$=92 K) on the cluster size $S$ (Fig. 5). Then, relying on the data on the statistics of finite NUC clusters as a function of the doping level δ (e.g., in YBCO), we can determine the dependences $T^*(\delta)$ and $T_c(\delta)$ [8], the result being in excellent agreement with the experiment.

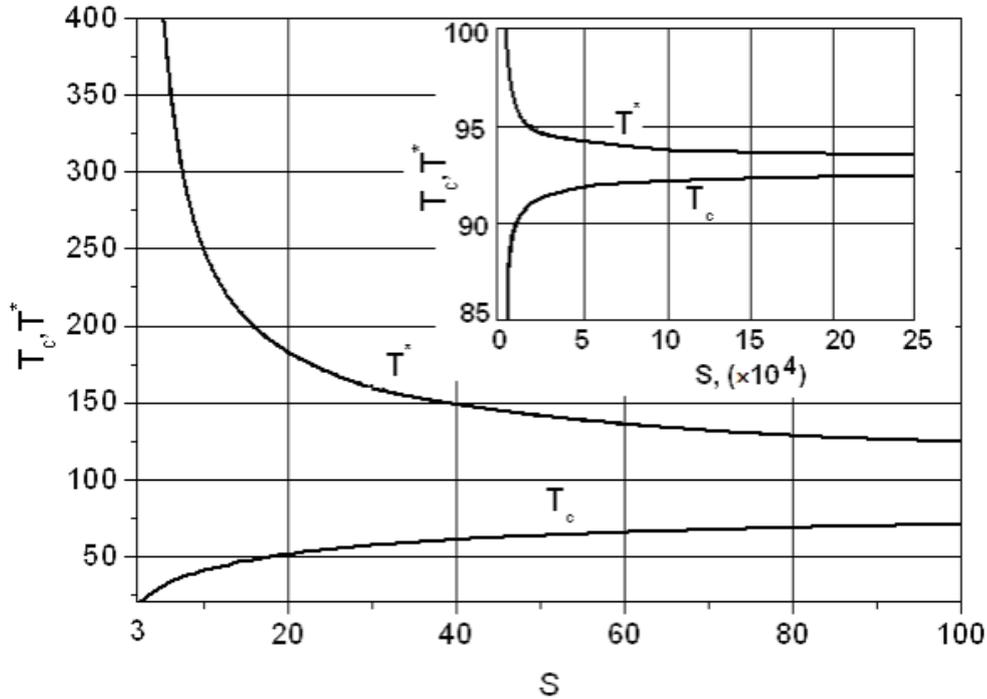

Fig. 5. Dependences of temperatures $T^*$ and $T_c$ for YBCO (with $T_{c\infty}$=92 K) on the cluster size $S$ for $S$<100. Inset: the same dependences for $S$<2.5×10$^5$.

Thus, in the region between curves $T_c(\delta)$ and $T^*(\delta)$, clusters fluctuate between the superconducting (coherent) and normal (incoherent) states. The number of NUC clusters being in the superconducting state at a given moment, as well as the lifetime of this state, increase with decreasing temperature. The experimentally measured value of $T_c$ has the meaning of temperature corresponding to the appearance of a percolation cluster of Josephson-coupled superconducting clusters of NUCs. It is evident, however, that, in a certain range of temperatures $T_c(\delta)<T<T^*(\delta)$, sufficiently long-lived and sufficiently large (although non-percolative) coherent superconducting clusters will be present. In these clusters, the Nernst effect and giant diamagnetism can be observed at $T>T_c$ [6,7]. The above discussion suggests that manifestation of these anomalies is not directly caused by the existence of the pseudogap, but rather results from the presence of fluctuating coherent superconducting clusters in the sample.

**Mechanism of carrier relaxation**

Because of the interaction of electrons with NUCs, the distribution of mobile holes is nondegenerate. Taking into account the absence of the degeneracy (i.e., the absence of the Pauli blocking) and high concentration of holes, the hole–hole scattering will be the dominant relaxation process.

However, as far as the interaction of two holes in a system with NUCs is effectively attractive, this process is not a usual Coulomb scattering. In this situation, the main scattering mechanism is similar to the one characteristic of metals and alloys with strong electron–phonon coupling [27]. In these materials, there exists an effectively attractive interaction between electrons occupying a layer of thickness $k\Theta_D$ (where $\Theta_D$ is the Debye temperature) at the FS, which originates from the exchange by virtual phonons and exceeds considerably the screened Coulomb repulsion. So, the main channel of electron–electron scattering in this case is also related to the virtual-phonon exchange. The contribution from these processes [27] becomes significant for $T<\Theta_D$. It proves that the amplitude of electron–electron scattering is independent of the energy $E$ of the colliding particles for $E<<k\Theta_D$ and drops abruptly when $E\sim k\Theta_D$. For $E>k\Theta_D$, only the Coulomb interaction contributes to the scattering amplitude. Experimentally, a contribution of the electron–electron scattering to the electrical resistivity $\rho$ ($\rho=AT^2$) exceeding the electron–phonon contribution was observed in Al [28] for $T<4$ K and in superconductors with A15 lattice [29] for $T<50$ K; the amplitude $A$ was found to be greater than the value calculated under the assumption of a purely Coulomb scattering mechanism by more than an order of magnitude.

Similarly, we suggest that the main contribution to the processes of hole carrier relaxation in HTSC is related to hole–hole scattering with the formation of an intermediate state bound at an NUC, which can be described as exchange by a virtual boson with energy $\Omega$. Inasmuch as $\Omega\sim\Delta E_U\sim 0.2$ eV, the range of temperatures where the contribution of this mechanism is significant extends to $T\sim 10^3$ K.

In this context, the temperature dependence of the resistivity can be obtained from the Drude formula $\rho=m^*\nu/n\cdot e^2$ (here, $m^*$ is the effective mass of holes and $\nu$ is the rate of hole scattering). For $\Omega>>E$, the scattering amplitude will be independent of the carrier energy $E$. For this reason, the scattering rate $\nu$ is determined by the concentration of holes and the statistical factor in the scattering cross section, i.e., the volume of the phase space available for the scattered particles. This factor is proportional to $E_1+E_2$ (where $E_1$ and $E_2$ are the energies of colliding particles counted from the chemical-potential level). Then,

$$\nu\propto n\cdot(E_1+E_2)$$

For the case of static conductivity, $E_1 \sim E_2 \sim \Gamma \propto T$, $\nu \propto nT \propto T^2/(T+T_0)$ and, thus, $\rho(T) \propto T$. It is this dependence that is observed experimentally for optimally doped samples of $YBa_2Cu_3O_7$, $La_{2-x}Sr_xCuO_4$, $Bi_2Sr_2CaCu_2O_y$ and other HTSCs.

In overdoped HTSCs, clusters of NUCs are immersed in a normal-metal matrix with a degenerate electron distribution. In this case, the resistivity term related to the hole–hole scattering mediated by NUCs assumes the form

$$\rho(T) \propto T^2/(T+T_0)$$

It is this dependence that is observed for various HTSC materials in the "overdoped" regime.

The fact that scattering processes are dominated by the electron–electron interaction will also reveal itself in the frequency and temperature dependences of the real part of the optical conductivity $\sigma_{opt}$:

$$\sigma_{opt} = (e^2 n/m^*) \cdot [\nu/(\omega^2 + \nu^2)]$$

(here, $\omega$ is the radiation frequency and $\nu$ is the "optical" relaxation rate). In the case of hole–hole scattering, the collision rate $\nu \geq 10^{15}$ s$^{-1}$ (for concentrations $n \sim 10^{22}$ cm$^{-3}$). This means that $\nu \gg \omega$ in the IR range, and the formula for the real part of the conductivity simplifies further to yield

$$\sigma_{opt} = e^2 n/m^* \nu$$

For the optical relaxation, $E_1 \sim \omega$, $E_2 \sim \Gamma \propto T$ and $\nu \propto n \cdot \omega$ for $\omega \gg \Gamma$, while $\nu \propto n \cdot T$ for $\omega \ll \Gamma$. Then, $\sigma_{opt} \propto \omega^{-1}$ for $\omega \gg \Gamma$ and $\sigma_{opt} \propto T^{-1}$ for $\omega \ll \Gamma$.

**Conclusions**

Thus, the suggested model provides a qualitative explanation for a number of anomalous properties of HTSCs According to this model, anomalies result from the unique electronic structure of HTSCs (a "Fermi-liquid insulator"), which favors the formation of two-atom NUCs and realization of unusual mechanism of electron–electron interaction.

**Acknowledgments**

This research was supported by Russian Foundation for Basic Research (grant 08-02-00881) and the Ministry of Education and Science of the Russian Federation (contract P2545).